\input epsf
\documentstyle[preprint,aps,tighten,epsf]{revtex}

\begin{document}

\draft

\title{
Quasi-Local Energy Conservation Law Derived \\
From The Einstein's Equations}

\author{ Jong Hyuk Yoon} 
\address{
Department of Physics\\
Kon-Kuk University, Seoul 143-701, Korea}

\maketitle
\thispagestyle{empty}

\begin{abstract}
The quasi-local energy conservation law is derived from 
the vacuum Einstein's equations on the timelike boundary surface 
in the canonical (2,2)-formalism of general relativity. 
The quasi-local energy and energy flux integral 
agree with the standard results in the asymptotically flat
limit and in spherically symmetric spacetimes. 
\end{abstract}


\newpage

In general relativity there have been many attempts to derive
quasi-local conservation laws\cite{brown,hayward1,hawking}. 
These conservation laws,
if they exist, would serve as useful selection rules
constraining the future development of a spacetime, 
given the fact that the direct integration of 
the Einstein's evolution equations is practically impossible. 
Recall that in the Newtonian theory, the conservation of momentum
\begin{math}
\sum\vec{p}={\rm const.}
\end{math} \ 
immediately follows from Newton's third law,
which is no more than the consistency condition implementing 
the second law. In general relativity, the consistency conditions 
for the evolution are already incorporated 
into the Einstein's equations via the constraint equations.
Therefore it has been long thought that 
conservation laws should follow directly
from the Einstein's equations\cite{derive,adm}.
The purpose of this paper is to show that 
this is indeed the case,
and using the canonical (2,2)-formalism of the vacuum
general relativity, 
I shall derive from the Einstein's equations on 
the {\it timelike} boundary surface 
the quasi-local energy conservation law,  
which relates the quasi-local gravitational energy loss 
to the net gravitational energy flux across a finite region 
of a given spacetime.

Let us start from the following line element\cite{unti}
\begin{equation}
ds^2 = -2dudv - 2hdu^2 +{\rm e}^{\sigma} \rho_{ab}
 \left( dy^a + A_{+}^{\ a}du +A_{-}^{\ a} dv \right)
\left( dy^b + A_{+}^{\ b}du +A_{-}^{\ b} dv 
\right), \label{yoon}
\end{equation}
where $+,-$ stands for $u,v$, respectively, and 
$\rho_{ab}(a,b=2,3)$ is the conformal 2-metric
of the transverse two-surface $N_{2}$ defined by 
$u,v ={\rm constant}$, which satisfies the condition 
\begin{equation}
{\rm det}\ \rho_{ab}=1.                        \label{det}
\end{equation}
Using the spacetime diffeomorphism invariance, 
one can always restrict 
the metric to the above form. 
Notice that these fields are 
functions of all the coordinates $(u,v,y^{a})$,
since we assume no isometries.
The integral $I_{0}$ of the scalar curvature of 
the metric (\ref{yoon}) is,
\begin{equation}
I_{0} =   \int \! \! du \, dv \, d^{2}y \, L_{0} 
+ {\rm surface}  \ {\rm integral},       \label{bareact}
\end{equation}
where
$L_{0}$ is given by\cite{alone}
\begin{eqnarray}
L_{0}& = &-{1\over 2}{\rm e}^{2 \sigma}\rho_{a b}
  F_{+-}^{\ \ a}F_{+-}^{\ \ b}
  +{\rm e}^{\sigma} (D_{+}\sigma) (D_{-}\sigma) 
  -{1\over 2}{\rm e}^{\sigma}\rho^{a b}\rho^{c d}
 (D_{+}\rho_{a c})(D_{-}\rho_{b d})  \nonumber\\
& & +{\rm e}^{\sigma} R_2
-2{\rm e}^{\sigma}(D_{-}h)(D_{-}\sigma) 
- h {\rm e}^{\sigma}(D_{-}\sigma)^2
+{1\over 2}h  {\rm e}^{\sigma}\rho^{a b}\rho^{c d}
 (D_{-}\rho_{a c})(D_{-}\rho_{b d}).     \label{barelag}
\end{eqnarray}
Here $R_{2}$ is the scalar curvature of $N_{2}$, and 
the notations are summarized below:
\begin{eqnarray}
& &F_{+-}^{\ \ a}=\partial_{+} A_{-} ^ { \ a}-\partial_{-}
  A_{+} ^ { \ a} - [A_{+}, A_{-}]_{\rm L}^{a}  \nonumber\\
& &\hspace{.95cm}=\partial_{+} A_{-} ^ { \ a}-\partial_{-}
  A_{+} ^ { \ a}-A_{+}^{\ c}\partial_{c}A_{-}^{\ a}
  +A_{-}^{\ c}\partial_{c}A_{+}^{\ a},   \label{field}\\
& &D_{\pm}\sigma = \partial_{\pm}\sigma
-[A_{\pm}, \sigma]_{\rm L}        \nonumber\\
& &\hspace{.95cm}=\partial_{\pm}\sigma
-A_{\pm}^{\ a}\partial_{a}\sigma
-\partial_{a}A_{\pm}^{\ a},         \label{get}\\
& &D_{\pm}h= \partial_{\pm}h - [A_{\pm}, h]_{\rm L} \nonumber\\
& &\hspace{.95cm}
=\partial_{\pm}h -A_{\pm}^{\ a}\partial_{a}h, \nonumber\\
& &D_{\pm}\rho_{a b}=\partial_{\pm}\rho_{a b}
   - [A_{\pm}, \rho]_{{\rm L}a b}       \nonumber\\
& &\hspace{1.25cm}=\partial_{\pm}\rho_{a b}
-A_{\pm} ^ { \ c}\partial_c \rho_{a b}
-(\partial_a A_{\pm} ^ { \ c})\rho_{c b}
-(\partial_b A_{\pm} ^ { \ c})\rho_{a c}
+(\partial_c A_{\pm} ^ { \ c})\rho_{a b},    \label{rhod}
\end{eqnarray}
where $[A_{\pm}, \ast]$ is the Lie derivative 
of $\ast$ along the vector field
$A_{\pm}:=A_{\pm}^{\ a}\partial_{a}$.
The integral $I_{0}$ appears as an action integral of 
a Yang-Mills type theory defined on
a fibre bundle, which consists of a (1+1)-dimensional base 
manifold coordinatized by $(u,v)$, 
and the 2-dimensional vertical space $N_{2}$ coordinatized 
by $y^{a}$.
In this picture, the Yang-Mills fields are valued in 
the Lie algebra associated with the group of 
the diffeomorphisms of $N_{2}$\cite{alone,four}.

In addition to the eight equations of motion that follow 
from (\ref{bareact}) by variations, there are 
two equations associated with gauge-fixing 
the spacetime metric to the form (\ref{yoon}). 
These equations, from which the quasi-local energy conservation law 
is shown to follow, are obtained by varying 
the Einstein-Hilbert action  before the gauge 
fixing condition (\ref{yoon}) is introduced\cite{four,dinsmall}.
They are found to be 
\begin{eqnarray}
&({\rm i})&\hspace{.5cm} {\rm e}^{\sigma} D_{+}D_{-}\sigma 
+ {\rm e}^{\sigma} D_{-}D_{+}\sigma  
+ 2{\rm e}^{\sigma} (D_{+}\sigma)(D_{-}\sigma) 
- 2{\rm e}^{\sigma}(D_{-}h)(D_{-}\sigma)  \nonumber\\
& &\hspace{.5cm} - {1\over 2}{\rm e}^{ 2 \sigma}\rho_{a b}
   F_{+-}^{\ \ a}F_{+-}^{\ \ b}
- {\rm e}^{\sigma} R_{2}      
- h {\rm e}^{\sigma} \Big\{
(D_{-}\sigma)^{2} 
-{1\over 2}\rho^{a b}\rho^{c d} 
 (D_{-}\rho_{a c})(D_{-}\rho_{b d})\Big\}=0, \label{ff}\\
&({\rm ii})&\hspace{.5cm} 
-{\rm e}^{\sigma} D_{+}^{2}\sigma 
- {1\over 2}{\rm e}^{\sigma}(D_{+}\sigma)^{2}
-{\rm e}^{\sigma}(D_{-}h) (D_{+}\sigma) 
+{\rm e}^{\sigma}(D_{+}h)(D_{-}\sigma) \nonumber\\
& &\hspace{.5cm}  +2h {\rm e}^{\sigma}(D_{-}h)(D_{-}\sigma) 
+{\rm e}^{\sigma}F_{+-}^{\ \ a}\partial_{a}h  
-{1\over 4}{\rm e}^{\sigma}\rho^{a b}\rho^{c d} 
 (D_{+}\rho_{a c})(D_{+}\rho_{b d})
+\partial_{a}\Big( \rho^{a b}\partial_{b}h \Big) \nonumber\\
& &\hspace{.5cm} 
+h {\rm e}^{\sigma}
\Big\{ - (D_{+}\sigma) (D_{-}\sigma)
+{1\over 2}\rho^{a b}\rho^{c d} (D_{+}\rho_{a c})
    (D_{-}\rho_{b d}) 
+{1\over 2}{\rm e}^{\sigma}\rho_{a b}F_{+-}^{\ \ a}F_{+-}^{\ \ b}
+R_{2} \Big\} \nonumber\\
& &\hspace{.5cm} 
+h^{2}{\rm e}^{\sigma}\Big\{
(D_{-}\sigma)^{2}
-{1\over 2}\rho^{a b}\rho^{c d}
(D_{-}\rho_{a c}) (D_{-}\rho_{b d})\Big\}=0.  \label{gg}
\end{eqnarray}
The above equations, being {\it first-order}
in $D_{-}$, should be regarded as constraint equations
associated with the partial gauge-fixing of  
the spacetime diffeomorphisms. Thus, in this formalism, 
the {\it natural} 
vector field with respect to which the evolution is to be defined 
is $D_{-}$. Let us define the momenta 
\begin{math}
\pi_{I}=\{ \pi_{h},\pi_{\sigma}, \pi_{a}, \pi^{a c} \} 
\end{math} \
conjugate to 
\begin{math}
q^{I}=\{ h, \sigma,   A_{+} ^ { \ a}, \rho_{a b} \}
\end{math} \
as
\begin{equation}
\pi_{I}:={\partial L_{0}\over \partial \dot{q}^{I}}, \label{momenta}
\end{equation}
where $\dot{q}^{I}:=D_{-}q^{I}$. Then we have
\begin{eqnarray}
& &\pi_{h}=-2 {\rm e}^{\sigma}(D_{-}\sigma),    \nonumber\\
& &\pi_{\sigma} = -2 {\rm e}^{\sigma} (D_{-}h)
       -2h {\rm e}^{\sigma} (D_{-}\sigma)
     + {\rm e}^{\sigma} (D_{+}\sigma), \nonumber\\
& &\pi_{a}={\rm e}^{2 \sigma} \rho_{a b}F_{+-}^{\ \ b}, \nonumber\\
& &\pi^{a c}= 
h{\rm e}^{\sigma} \rho^{a b}\rho^{c d}(D_{-}\rho_{b d})
-{1\over 2}{\rm e}^{\sigma} \rho^{a b}\rho^{c d}
(D_{+}\rho_{b d}),                            \label{first}
\end{eqnarray}
where $\pi^{a c}$ is traceless since  
\begin{math}
{\rm det} \rho_{ab}=1.
\end{math}
The Hamiltonian density
$H_{0}$ defined as 
\begin{equation}
H_{0}:=\pi_{I}\dot{q}^{I} - L_{0}
\end{equation}
is given by 
\begin{eqnarray}
& &H_{0} =  -{1\over 2}{\rm e}^{-\sigma}\pi_{\sigma}\pi_{h} 
+ {1\over 4}h{\rm e}^{-\sigma}\pi_{h}^{2} 
-{1\over 2}{\rm e}^{-2\sigma}\rho^{a b}\pi_{a}\pi_{b}
+{1\over 2h}{\rm e}^{-\sigma}\rho_{a c}\rho_{b d}\pi^{a b}\pi^{c d}
+{1\over 2}\pi_{h}(D_{+}\sigma)   \nonumber\\
& & \hspace{1.1cm}+{1\over 2h}\pi^{a b}(D_{+}\rho_{a b}) 
+{1\over 8h}{\rm e}^{\sigma}\rho^{a b} \rho^{c d}
(D_{+}\rho_{a c}) (D_{+}\rho_{b d}) 
-{\rm e}^{\sigma}R_{2}.          \label{tilde}
\end{eqnarray}
In terms of these variables
the two equations (\ref{ff}) and (\ref{gg}) become 
\begin{eqnarray}
&({\rm i}) \hspace{.3cm} & H_{0} - \partial_{+}\pi_{h} 
+ \partial_{a} \Big( 
A_{+}^{\ a}\pi_{h} 
+ {\rm e}^{-\sigma}\rho^{a b} 
\pi_{b} \Big)=0,                    \label{qmomentum}\\
&({\rm ii}) \hspace{.3cm} & \pi_{h}D_{+}h + \pi_{\sigma}D_{+}\sigma 
+  \pi^{a b}D_{+}\rho_{a b}     
-\partial_{+}\Big( 
     2h\, \pi_{h} + 2 {\rm e}^{\sigma}D_{+}\sigma \Big)   
+\partial_{a}\Big( 
    2h\, \pi_{h}A_{+}^{ \ a}      \nonumber\\
&  \hspace{.3cm}&
+ 2A_{+}^{ \ a}{\rm e}^{\sigma}D_{+}\sigma 
 + 2h {\rm e}^{-\sigma}\rho^{a b}\pi_{b}  
 +2\rho^{a b}\partial_{b}h \Big) =0.         \label{qenergy}
\end{eqnarray}
It is from the equation (\ref{qenergy}) that the quasi-local 
energy conservation law follows. 
To see this, let us consider the asymptotically flat spacetimes 
which approach the Schwarzschild solution 
\begin{equation}
ds^2 \longrightarrow 
-2dudv-\Big( 1-{2m\over v}\Big)\, du^2
 + v^{2} d\Omega^{2}        \label{sol}
\end{equation}
in the asymptotic limit. Here 
\begin{math}
\partial / \partial u
\end{math} \ 
is the timelike Killing vector field.
Let us suppose that the spacetime we are considering is
asymptotically flat, so that $N_{2}$ is a compact two-surface 
without boundary which is asymptotic to $S_{2}$, 
and that the $u$-coordinate is asymptotic 
to the Killing time (see Figure 1). From the metric (\ref{yoon}),
we find that the hypersurface defined 
by $du=0$ is a null hypersurface, 
and the hypersurface defined by $dv=0$ is either a timelike or
a spacelike hypersurface. Let us restrict our discussion to 
the region where $dv=0$ is a timelike hypersurface.
The integral of the equation (\ref{qenergy}) 
over a timelike hypersurface
\begin{math}
N_{2}\times [u_{0}, u_{1}]
\end{math} \ 
becomes, with a suitable normalization,
\begin{eqnarray}
& &{1\over 16\pi}\oint_{ u_{1}} 
    \! \! \! d^{2}y \, \Big( 
     h\, \pi_{h} +  {\rm e}^{\sigma}D_{+}\sigma \Big)   
- {1\over 16\pi}\oint_{ u_{0}}  
    \! \! \!  d^{2}y \, \Big( 
   h\, \pi_{h} +  {\rm e}^{\sigma}D_{+}\sigma \Big) \nonumber\\
& &  \hspace{5cm}  
={1\over 32\pi}\int_{u_{0}}^{u_{1}} \! \! \! \! \! du \!
  \oint_{u} \! d^{2}y \, \Big(
   \pi_{h}D_{+}h +\pi_{\sigma}D_{+}\sigma  
  + \pi^{a b}D_{+}\rho_{a b}\Big).            \label{enflux}
\end{eqnarray}

In general, the energy flux across a surface is given by
the energy momentum tensor $T_{0i}$, which is of the form
\begin{equation}
T_{0i}\sim \pi_{\phi}\partial_{i}\phi
\end{equation}
in scalar field theories, 
where $\pi_{\phi}$ is the conjugate momentum of 
the scalar field $\phi$.
The integrand of the r.h.s. of (\ref{enflux}) is precisely of this
form, and therefore it represents 
the gravitational energy flux across 
the two-surface $N_{2}$, integrated from $u_{0}$ to $u_{1}$.
Then the l.h.s. of (\ref{enflux}) should represent
the net change of the gravitational energy 
of the region enclosed by the compact two-surface $N_{2}$. 
Thus our proposal of the quasi-local gravitational 
energy $E(u,v)$ of a region enclosed by a two-surface $N_{2}$ 
on the timelike surface defined by $v={\rm constant}$ 
at the instant $u$ is 
\begin{equation}
E(u,v):= {1 \over 16\pi}\oint_{u} d^{2}y \,   \Big( 
h\, \pi_{h} + {\rm e}^{\sigma} D_{+}\sigma \Big) 
+ E_{0}(v),              \label{casi}
\end{equation}
where $E_{0}(v)$ is an undetermined substraction term
which is also a two-surface integral.
The equation (\ref{enflux}) then becomes
\begin{equation}
E(u_{1},v)-E(u_{0},v)
={1 \over 32\pi} \int_{u_{0}}^{u_{1}} \! \!  \! \! \!  du \!
  \oint_{u} \! d^{2}y \, \Big(
   \pi_{h}D_{+}h + \pi_{\sigma}D_{+}\sigma 
  + \pi^{a b}D_{+}\rho_{a b}\Big).           \label{enflux2}
\end{equation}
The flux integral in general does not have a definite sign, since
it includes the flux carried by the in-coming as well as 
the out-going gravitational waves. 
But in the asymptotically flat region, 
the flux integral turns out to be negative-definite, 
representing the physical situation that 
there is no in-coming flux coming from the infinity.

For the asymptotically flat spacetimes, let us show that
the relation (\ref{enflux2}) reduces to the standard Bondi 
mass-flux relation\cite{tod} in the asymptotic limit. 
If we choose the substraction term $E_{0}(\infty)$ at infinity as
\begin{equation}
E_{0}(\infty)= \lim_{v\rightarrow \infty}{1 \over 16\pi}\oint 
 \! d \Omega \, 
\Big( v^{2}\partial_{-}{\sigma} \Big),      \label{ezero}
\end{equation}
the total energy $E(u,\infty)$ becomes 
\begin{equation}
E(u,\infty)=m(u),
\end{equation}
where $m(u)$ is the Bondi mass of the asymptotically flat 
radiating spacetime.
In order to show that the r.h.s. of (\ref{enflux2}) becomes 
the corresponding Bondi flux integral, we need to 
know the asymptotic 
fall-off rates of the fields in a more detail. 
In the asymptotic region where an out-going twist-free 
null vector field exists, the metric is approximated as
\begin{eqnarray}
& & ds^2 \longrightarrow 
-2dudv - 2hdu^2 +{\rm e}^{\sigma} \rho_{ab}
    \left( dy^a + A_{+}^{\ a}du \right) 
 \left( dy^b + A_{+}^{\ b}du  \right), \label{yoona}
\end{eqnarray}
where $\partial / \partial v$ is the twist-free 
null vector field. If we compare this line element 
with the Schwarzschild spacetime (\ref{sol}), 
then the fall-off rates of the fields are found to be
\begin{eqnarray}
& &{\rm e}^{\sigma}=v^{2}   ({\rm sin}\vartheta) \Big\{ 
1+ {\alpha (u,\vartheta,\varphi)\over v} 
+O({1 / v^{2}})  \Big\},           \nonumber\\
& &\rho_{\vartheta \vartheta}
=\Big( {1\over {\rm sin}\vartheta} \Big) \Big\{ 
1 + {\beta (u,\vartheta,\varphi)\over v}
+ O({1 / v^{2}}) \Big\},        \nonumber\\  
& &\rho_{\varphi \varphi}=({\rm sin}\vartheta)  \Big\{ 
    1 + {\gamma (u,\vartheta,\varphi)\over v} 
+ O({1 / v^{2}}) \Big\},     \nonumber\\
& &\rho_{\vartheta \varphi}
={\delta (u,\vartheta,\varphi)\over v}+ O({1 /  v^{2}}),\nonumber\\
& &2h=1-{2m\over v}+ O({1 / v^{2}}),   \nonumber\\
& &A_{+}^{\ a}=O({1 / v}),  \label{asymp}
\end{eqnarray}
where $\alpha, \beta, \gamma,$ and $\delta$ are arbitrary functions of 
$(u,\vartheta,\varphi)$. 
Using these conditions, the r.h.s. of (\ref{enflux2}) becomes, 
\begin{eqnarray} 
& & {1\over 32\pi} \int_{u_{0}}^{u_{1}} \! \!\! \! \! du \!
  \oint_{u} \! d\vartheta d\varphi \, \Big(
\pi_{h}D_{+}h + \pi_{\sigma}D_{+}\sigma  
+ \pi^{a b}D_{+}\rho_{a b}      \Big)           \nonumber\\ 
& &={1\over 32\pi}\int_{u_{0}}^{u_{1}}\! \! \! \! \! du \!
  \oint_{u} \! d\vartheta d\varphi \, \Big\{
 -{1\over 2}{\rm e}^{\sigma}\rho^{a b}\rho^{c d}
(\partial_{+}\rho_{a c})(\partial_{+}\rho_{b d})   
+{\rm e}^{\sigma} (\partial_{+}\sigma)^{2}
-2h {\rm e}^{\sigma} (\partial_{+}\sigma)  
(\partial_{-}\sigma) \Big\}. \label{almost}
\end{eqnarray}
As is shown below, 
the integral (\ref{almost}) can be made negative-definite, 
if we use the residual symmetry\cite{tod} 
of the metric (\ref{yoona}) which consists of, 
apart from the obvious diffeomorphisms of $N_{2}$,
the following coordinate transformation
\begin{equation}
v \longrightarrow 
v'=v + \epsilon (u,\vartheta,\varphi),  \label{rescue}
\end{equation}
while keeping ($u,\vartheta,\varphi$) constant.
Here $\epsilon$ is an {\it arbitrary} function of 
$(u,\vartheta,\varphi)$.
This symmetry transformation reflects the freedom of 
choosing the origin of the affine parameter $v$ 
in an arbitrary way. In the new coordinate system, we
find from (\ref{asymp}) that the asymptotic form of 
${\rm e}^{\sigma}$ becomes
\begin{equation}
{\rm e}^{\sigma}
= v'^{2}({\rm sin}\vartheta) \Big\{ 
1+{1\over v'}( \alpha -2\epsilon )
+O({1 / v'^{2}})         \Big\}.        \label{aream}
\end{equation}
Let us choose the function $\epsilon$
\begin{equation}
\epsilon={1\over 2}\alpha,
\end{equation}
such that ${\rm e}^{\sigma}$ becomes
\begin{equation}
{\rm e}^{\sigma}= v'^{2}({\rm sin}\vartheta ) \Big\{ 
1 + 0(1/v'^{2}) \Big\}.              \label{absent}
\end{equation}
Then we have
\begin{equation}
\partial_{+}\sigma = 0(1/v'^{2}), \hspace{1cm}
\partial_{-}\sigma = 0(1/v'),                \label{neworder}
\end{equation}
so that the last two terms in the r.h.s of (\ref{almost}) 
vanish in the limit $v' \rightarrow \infty$. 
Therefore the energy difference (\ref{enflux2}) becomes
\begin{equation}
E(u_{1},\infty) -  E(u_{0},\infty)  
=-{1\over 64\pi}\int_{u_{0}}^{u_{1}} \! \!\!\! \! du \!
  \oint_{u} \! d\Omega \, v^{2}
  \rho^{a b}\rho^{c d}
(\partial_{+}\rho_{a c})(\partial_{+}\rho_{b d}) 
\leq 0.                            \label{lindo}
\end{equation} 
This relation is precisely the Bond mass-loss formula, relating 
the mass loss in terms of a negative-definite flux 
of the out-going gravitational waves. The r.h.s. of the integral 
is a bilinear of the gravitational {\it current} $j^{a}_{ \ b}$
\begin{equation}
j^{a}_{ \ b}=\rho^{a c}\partial_{+}\rho_{ b c},
\end{equation}
which is precisely the shear degrees of 
freedom of the gravitational field.

Let us now consider a spherical ball
of radius $v$, filled with a perfect fluid with energy density 
$\rho(v)$, and find the total energy contained in the ball. 
The Einstein's equations are modified by the presence of 
the fluid, and the solution is given by 
\begin{equation}
ds^2 = -2dudv - 2h(v)du^2 + v^{2} 
    d \Omega ^{2},             \label{fluid}
\end{equation}
where 
\begin{equation}
2h(v)=1-{2m(v)\over v},  \hspace{0.5cm}
m(v)=4\pi \! \! \int_{0}^{v} \! \! 
d v' v'^{2}\, \rho(v').      \label{twoei}
\end{equation}
The substraction term at the infinity $E_{0}(\infty)$ 
in (\ref{ezero}) suggests that the substraction term 
$E_{0}(v)$ at a finite distance should be chosen as
\begin{equation}
E_{0}(v)={1 \over 16\pi}\oint_{ S_{2}}
\! \! \! d \Omega
\Big( v^{2} \partial_{-}\sigma  \Big).           \label{substr}
\end{equation}
Then $E(u,v)$ becomes
\begin{eqnarray}
E(u,v) & = & {1 \over 16\pi}\oint_{S_{2}}
\! \! \! d \Omega
\Big(  1-2h \Big) \Big( v^{2} \partial_{-}\sigma \Big)  \nonumber\\
& = & m(v),       \label{asycorrecto}
\end{eqnarray}
which is identical to the Misner-Sharp energy\cite{misharp}
for spherically symmetric distribution of an ideal fluid.
However, this is {\it not} the proper mass. 
The proper volume element $dV$ of the $u={\rm constant}$
null hypersurface is given by
\begin{equation}
(2h)^{-1/2}v^{2}dv d\Omega ,   \label{deternull}
\end{equation}
so that the proper quasi-local mass $m_{\rm p}(v)$ is given by
\begin{eqnarray}
m_{\rm p}(v)&=&4\pi \! \! \int_{0}^{v} \! \! 
d v' v'^{2}\, (2h)^{-1/2} \rho(v')  \nonumber\\
&=& 4\pi \! \! \int_{0}^{v} \! \! 
d v' v'^{2}\, \Big( 
1-{2m(v')\over v'}\Big)^{-1/2}\rho(v').  \label{twoeia}
\end{eqnarray}
The difference between  $E(u,v)$ and $m_{\rm p}(v)$ can be
interpreted as the gravitational binding energy 
$E_{\rm B}(v)$, so that $E(u,v)$ becomes
\begin{equation}
E(u,v)=m_{\rm p}(v) - E_{\rm B}(v) .   \label{bindinga}
\end{equation}
In the Newtonian limit where $E(u,v)/v$ is small, 
$E_{\rm B}(v)$ becomes
\begin{equation}
E_{\rm B}(v)
\approx  4\pi \! \! \int_{0}^{v} \! \! 
d v' v'^{2}\, \rho(v')  
\Big( {m_{\rm p}(v')\over v'}\Big),   \label{nlimit}
\end{equation}
which is just the total Newtonian binding energy of $m_{\rm p}(v')$ 
interacting with the fluid shell of mass 
\begin{math}
4\pi d v' v'^{2}\, \rho(v'),
\end{math} \ 
integrated over the ball. 
Notice that $E_{\rm B}(v)$ has the correct sign\cite{brown},
so $E(u,v)$ indeed represents the total gravitational energy 
within the spherical ball of the radius $v$, 
including the binding energy. 

It is worth mentioning that from the equation 
(\ref{qmomentum}) the quasi-local conservation
of {\it momentum} also follows.
The resulting
equation has a similar structure with the Euler equation
in fluid dynamics\cite{landau}. The Hamiltonian density
$H_{0}$ in this (2,2)-formalism, which corresponds to the 
{\it momentum flux}, consists of terms which are at most quadratic 
in $\pi_{I}$. From this point of view, terms linear 
in $\pi_{I}$ may be regarded as viscosity terms, 
and terms independent of $\pi_{I}$ as pressure terms.

In summary I derived
a quasi-local version of the mass-flux relation
for the {\it vacuum} general relativity 
from the Einstein's equation on the timelike 
boundary hypersurface.
The proposed quasi-local mass and flux integral 
are shown to reproduce 
the standard results in the asymptotic region, and agrees with
the Misner-Sharp energy for the spherically symmetric fluid 
in a finite region. Unlike other proposals of quasi-local 
gravitational energies, our expression is {\it canonical},
being derived from the first-order (2,2)-formalism. 
From the viewpoint of the correspondence principle, 
our derivation is particularly satisfactory, 
since the quasi-local conservation 
in this formalism comes from the ``consistency equation'', 
as in the Newtonian theory. 

It is a great pleasure to thank Prof. Joohan Lee for several 
enlightening discussions, and Prof. Carlos Kozameh 
for encouragement.
This work is supported by KOSEF, Grant No. 95-0702-04-01-3.

\nopagebreak

\newpage

\vspace{-6cm}
\begin{figure}
\epsfxsize=15cm \epsfbox{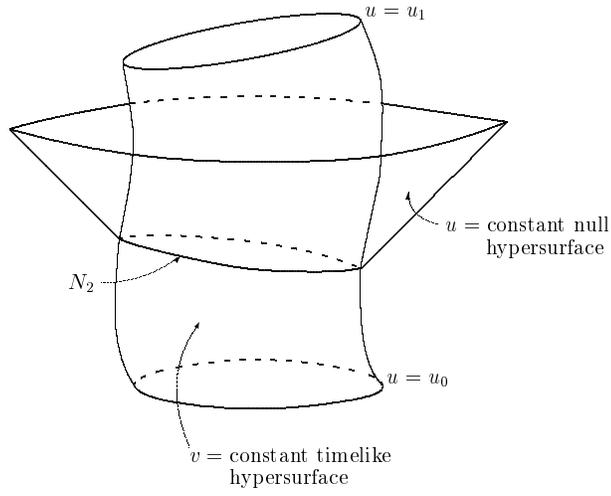}
\vspace{-7cm}
\caption{A region of spacetime where the surface $du=0$ 
is a null hypersurface, and the surface $dv=0$ 
is a timelike hypersurface. Their intersection defines 
$N_{2}$.
The vector field $\partial / \partial u$ is normal
to $N_{2}$ if and only if $A_{+}^{\ a}=0$, and its norm 
approaches $-1$ near the infinity.} 
\label{fig1}
\end{figure}


\begin{thebibliography}{99}
\bibitem[*]{email}{Electronic address:yoonjh@cosmic.konkuk.ac.kr}
\bibitem{brown}{J. Brown and J. York, Phys. Rev. D {\bf 47}, 
  1407 (1993).}
\bibitem{hayward1}{S. Hayward, Phys.Rev. D {\bf 49}, 
  831 (1994).}
\bibitem{hawking}{S. Hawking and C. Hunter,Class. Quant. Grav.
{\bf 13}, 2735 (1996).}
\bibitem{derive}{J.N. Goldberg, Phys. Rev. {\bf 111}, 315 (1958).}
\bibitem{adm}{R. Arnowitt, S. Deser, and C. Misner, 
  Phys. Rev. {\bf 122}, 997 (1961).}
\bibitem{unti}{E.T. Newman and T. Unti, J. Math. Phys. {\bf 3}, 
   891 (1992).}
\bibitem{alone}{J.H. Yoon, Phys. Lett. B {\bf 308}, 240 (1993).}
\bibitem{four}{Y.M. Cho, Q.H. Park, K.S. Soh, and J.H. Yoon, 
  Phys. Lett. B {\bf 286}, 251 (1992). }
\bibitem{dinsmall}{R. d'Inverno and J. Smallwood, 
 Phys. Rev. D {\bf 22}, 1233 (1980)}.
\bibitem{tod}{E.T. Newman and K.P. Tod, 
{\it Asymptotically Flat Space-Times}, in 
General Relativity and Gravitation, Vol.2, ed. A. Held, 
Plenum Press (1980) }.
\bibitem{misharp}{C. Misner and D. Sharp, Phys. Rev. {\bf 136}, 
  B571 (1964)}
\bibitem{landau}{L.D. Landau and E.M. Lifshitz, 
{\it Fluid Mechanics}, Pergamon Press (1989).}
\end{thebibliography}
\end{document}